\documentclass[12pt,preprint]{aastex}
\usepackage{natbib}
\begin{document}
\title{Far Ultraviolet Spectroscopy of Seven Nova-Like
Variables}

\author{Trisha Mizusawa}
\affil{Dept. of Astronomy and Astrophysics,
Villanova University,
Villanova, PA 19085,
email: trisha.Mizusawa@villanova.edu}

\author{Jason Merritt}
\affil{Dept. of Astronomy and Astrophysics,
Villanova University,
Villanova, PA 19085,
email: jason.meritt@villanova.edu}

\author{Ronald-Louis Ballouz}
\affil{Dept. of Astronomy and Astrophysics,
Villanova University,
Villanova, PA 19085,
email: ronald-louis.ballouz@villanova.edu}

\author{Michael Bonaro}
\affil{Dept. of Astronomy and Astrophysics,
Villanova University,
Villanova, PA 19085,
email: michael.bonaro@villanova.edu}

\author{Sean Foran}
\affil{Dept. of Astronomy and Astrophysics,
Villanova University,
Villanova, PA 19085,
email: sean.foran@villanova.edu}

\author{Christopher Plumberg}
\affil{Dept. of Astronomy and Astrophysics,
Villanova University,
Villanova, PA 19085, 
and 
Eastern University, Dept. of Physical Science, 
St.Davids, PA 19087,
email: astrophysicist87@gmail.com}

\author{Heather Stewart}
\affil{Dept. of Astronomy and Astrophysics,
Villanova University,
Villanova, PA 19085,
email: heather.stewart@villanova.edu}

\author{Trayer Wiley}
\affil{Dept. of Astronomy and Astrophysics,
Villanova University,
Villanova, PA 19085,
and 
Eastern University, Dept. of Physical Science, 
St.Davids, PA 19087,
email: trayer3@gmail.com}

\author{Edward M. Sion}
\affil{Dept. of Astronomy \& Astrophysics,
Villanova University,
Villanova, PA 19085, 
e-mail: edward.sion@villanova.edu}

\begin{abstract}
We present the results of a multi-component synthetic spectral analysis of 
the archival far ultraviolet spectra of several key nova-like variables 
including members of the SW Sex, RW Tri, UX UMa and VY Scl subclasses: KR 
Aur, RW Tri, V825 Her, V795 Her, BP Lyn, V425 Cas and HL Aqr. Accretion 
rates as well as the possible flux contribution of the accreting white 
dwarf are included in our analysis. Except for RW Tri which has a reliable 
trigonometric parallax, we computed the distances to the nova-like systems 
using the method of \citet{Knigge06}. Our analysis of seven archival IUE 
spectra of RW Tri at its parallax distance of 341 pc consistently 
indicates a low mass ($\sim0.4$M$_{\sun}$) white dwarf and an average 
accretion rate, \.{M}$ = 6.3 \times 10^{-9}$ M$_{\sun}$/yr. For KR Aur, 
we estimate that the white dwarf has T$_{eff} = 29,000\pm2000$K, 
log $g = 8.4$ and contributes 18\% of the FUV flux while an accretion
disk with accretion 
rate \.{M}$ = 3 \times 10^{-10}$ M$_{\sun}$/yr at an inclination of 41 
degrees, contributes the remainder. We find that an accretion disk 
dominates the far UV spectrum of V425 Cas but a white dwarf contributes 
non-negligibly with approximately 18\% of the FUV flux. For the two high 
state nova-likes, HL Aqr and V825 Her, their accretion disks totally 
dominate with \.{M}$ = 1 \times 10^{-9}$ M$_{\sun}$/yr and $3\times 
10^{-9}$ M$_{\sun}$/yr, respectively. For BP Lyn we find \.{M}$ = 1 \times 
10^{-8}$ M$_{\sun}$/yr while for V795 Her, we find an accretion rate of 
\.{M}$ = 1 \times 10^{-10}$ M$_{\sun}$/yr. We discuss the implications of 
our results for the evolutionary status of nova-like variables.

\end{abstract}

Subject Headings: Stars: cataclysmic variables, white dwarfs, Physical
Processes: accretion, accretion disks

\section{Introduction}

Non-magnetic Cataclysmic variables (CVs) are short-period binaries in 
which a late-type, Roche-lobe-filling main-sequence dwarf transfers gas 
onto a rotating, accretion-heated white dwarf (WD). In a non-magnetic 
system, the white dwarf does not have a magnetic field strong enough to 
prevent the formation of an accretion disk. Hence, the gas carrying 
angular momentum is expected to accrete preferentially onto the equatorial 
latitudes of the white dwarf and spread rapidly with latitude 
\citet{Piro04,Balsara09}. For accretion rates 
below a critical threshold when the accretion flow would be so high (hot) 
that the accretion disk would become fully ionized 
\citep{Shafter86}. This critical threshold \citep{Cannizzo82,Cannizzo84} is given by

\.{M}crit$ = 1.38\times10^{16} r_{10}^{2.6} M_{1}^{-0.87}$ g/s. 

For a dwarf nova, a thermal-viscous instability in the disk leads to 
repetitive (limit cycle-like) disk accretion events in which gravitational 
energy is released (the dwarf nova outburst) in a few days followed by 
weeks to months of lower level accretion (dwarf nova quiescence) before 
the disk builds up to trigger the next outburst. If a cataclysmic variable 
white dwarf accretes enough envelope mass, then every few thousand years 
to tens of thousands of years (depending upon the white dwarf mass and 
average accretion rate), a thermonuclear explosion occurs in the white 
dwarf's accreted envelope which is identified as the classical nova 
outburst.

However, there is a subclass of CVs, known as the nova-like 
variables, in which the mass-transfer rate tends to be higher and the 
light of the system is typically dominated by a very bright accretion disk 
(Warner 1995). The spectra of nova-like variables generally resemble those 
of classical novae (CNe) that have settled back to quiescence. Yet the 
nova-like variables have never had a recorded CN outburst, dwarf nova 
outburst or any outburst. Hence their evolutionary status is still unclear. 
They could be close to having their next CN explosion, or they may have 
had an unrecorded explosion, in the recent past.  Their distribution of 
orbital periods covers a broad range above the upper boundary
of the CV period gap. Within the CV period gap between orbital periods 
of two and three hours, very few CVs are found.

The nova-like variables comprise a number of 
subgroups with differing photometric and spectroscopic behavior. The 
various subclasses of nova-like variables are defined in \citet{Warner95}. 
Some nova-likes (classified as the VY Sculptoris systems) exhibit the 
behavior of being in a high optical brightness state for most of the time, 
but unpredictably plummet into a deep low optical brightness state with 
little or no ongoing accretion. Then, just as unpredictably, their optical 
brightness returns to the high state (cf. \citealt{Kafka04} and 
references therein).

These precipitous drops in brightness are possibly related to the 
cessation of mass transfer from the K-M dwarf secondary star either by 
starspots that drift into position under the inner Lagrangian point, L1 
\citep{Livio98} or irradiation feedback in which an inflated outer 
disk can modulate the mass transfer from the secondary by blocking its 
irradiation by the hot inner accretion disk region \citep{Wu95}. Other 
nova-like systems, the UX UMa subclass, do not appear to exhibit low 
states but remain in a state of high accretion, sometimes referred to as 
dwarf novae stuck in permanent outburst (Warner 1995). It is widely 
assumed that the absence of dwarf novae outbursts in nova-likes is 
explained by their mass transfer rates being above the critical threshold 
where accretion rates are so high that the accretion disk is largely 
ionized, thus suppressing the viscous-thermal instability (the disk 
instability mechanism or DIM) which drives dwarf nova limit cycles 
\citep{Shafter86}. However, in recent years, a number of 
authors (e.g., \citealp{Borges05,Baptista07}) have 
presented evidence questioning the validity of the disk 
instability mechanism.

Still another nova-like subclass, this one spectroscopically-defined, is 
the SW Sextantis stars, which display a multitude of observational 
characteristics: orbital periods between 3 and 4 hours, one third of the 
systems non-eclipsing and two-thirds showing deep eclipses of the WD by 
the secondary, single-peaked emission lines despite the high inclination, 
and high excitation spectral features including He II (4686) emission and 
strong Balmer emission on a blue continuum, high velocity emission S-waves 
with maximum blueshift near phase $\sim 0.5$, delay of emission line 
radial velocities relative to the motion of the WD, and central absorption 
dips in the emission lines around phase $\sim 0.4 - 0.7$ \citep{Rodr07a}.  
The white dwarfs in 
many, if not all, of these systems are suspected of being magnetic 
\citep{Rodr07}. Since these objects are found near the upper 
boundary of the period gap, their study is of critical importance to 
understanding CV evolution as they enter the period gap \citep{Rodr07}.

The time-averaged accretion rates of all subtypes of nova-like variables
must be known in order to understand their secular evolution and to
compare their accretion rates with each other and with the various classes
of dwarf novae. These will in turn shed light on the evolutionary
properties, including the possible relation to supersoft sources and Type
Ia supernovae. In this paper, we use synthetic spectra of accretion disks
and white dwarf photospheres to carry out analyses of the previously
unmodeled far ultraviolet spectra of seven nova-like variables with a
special focus on determining their accretion rates. Two of the
best-observed nova-likes, KR Aur and RW Tri, each have relatively good
quality FUV coverage. RW Tri in particular also has an accurate
trigonometric parallax. These two objects together with V825 Her, V795
Her, BP Lyn, V425 Cas and HL Aqr represent three different subclasses of
nova-like systems. Four of the systems, BP Lyn, V795 Her, HL Aqr and RW
Tri, are also classified as SW Sextantis systems, while HL Aqr and RW Tri
are also classified as UX UMa systems. Two of the nova-like systems, KR
Aur and V425 Cas, are classified as VY Sculptoris systems. We briefly
describe their presently known overall properties.

\subsection{KR Aur}

This nova-like system is classified as a VY Scl subtype with high states 
at magnitude 12.7 and the deepest low state being 17.9 (\citealt{Ritter03}, Edition 7.12). The system 
underwent a deep minimum during 1994-95 when optical spectra revealed 
emission lines possibly indicative of a disk still being present \citep{Antov96}. The minimum lasted eight months. However, usually the system 
brightness is magnitude 12 to 14 with 13.5 being the most typical 
brightness. KR Aur is one of six out of 23 VY Scl nova-likes which have 
negative superhumps. Negative superhumps may arise when a disk becomes 
tilted due to intense radiation from a very hot white dwarf and thus could 
be an indirect indicator of nuclear burning \citep{Koz07}. The other 
systems with negative superhumps are V442 Oph, DW UMa, TT Ari and V751 
Cyg. The orbital inclination is low with an upper limit of 38 degrees and 
a lower limit of 10 degrees \citep{Ritter03}. The orbital period is 
0.1628 days. The white dwarf masss has been estimated to be in the range 
of 0.59 to 0.17 M$_{\sun}$ \citep{Ritter03}.

\subsection{RW Tri}

This very bright eclipsing nova-like variable ($V = 12.5$) is classified 
as both an SW Sex star and a UX UMa system since it never been observed to 
go into a low brightness state. It has an accurate Hubble FGS 
trigonometric parallax of $341\frac[+38\pm41]$ pc \citep{McArthur99}, and a 
moderately high orbital inclination of $70.5\pm2.5$ degrees \citep{Smak95a}. 
The component masses given in (\citet{Ritter03}; update 7.12) are 
M$_{wd} = 0.45\pm 0.15$ M$_{\sun}$ and a secondary mass of $0.63\pm0.1$ 
M$_{\sun}$ while Poole et al.(2003) found M$_{wd} = 0.55 \pm 0.15$ 
M$_{\sun}$ and M$_{2} = 0.35\pm0.05$ M$_{\sun}$.

\subsection{V825 Her}

This thick disk object, discovered as PG 1717+413, is a luminous, 
nova-like CV classified as a UX UMa-type nova-like variable \citep{Ferguson84, Ringwald05}, as though 
it were a dwarf nova stuck in outburst continuously. There is no evidence 
of coherent optical pulsations expected for a magnetic accretor and only 
very weak He II, whereas He II is typically strong in magnetic CVs. No physical 
parameters were previously determined for V825 Her \citep{Ritter03}. 
Its white dwarf mass, orbital inclination and accretion rate are unknown. 
\citet{Ringwald05} reported no spectroscopic trace of the secondary 
star in their optical spectra thus suggesting relatively high luminosity 
and mass transfer rate.

\subsection{V795 Her}

V795 Her is an SW Sex-subtype nova-like variable \citep{Casares96} which has an uncertain classification as an intermediate 
polar (it shows a coherent X-ray period which could be from a non-synchronously 
spinning, magnetized white dwarf and is a strong X-ray source). It exhibits a highly variable line
spectrum with a periodicity seen in IUE spectra (4.8 hours) that was mistakenly 
identified with its orbital period and a continuum steeply rising toward short wavelengths. The correct
orbital period appears to be 2.6 hours.  \citep{Rosen98} found the 2.6 hour 
orbital signature in the UV lines detected on HST FOS time series spectra. Nevertheless, there is still
some uncertainty associated with the origin of both detected periodicities. No eclipses 
have been observed.  There is possibly a P Cygni profile at C IV.

\subsection{BP Lyn}

BP Lyn (PG0859+4150) is a low inclination SW Sex-type nova-like variable 
where the white dwarf primary is not eclipsed by the secondary star (Hoard 
\& Szkody 1996). There is a shallow V-shaped eclipse suggesting an eclipse 
of a hot spot in the system around 0.9 phase. There are high excitation 
emission lines and the high Balmer lines have broad absorption wings, 
characteristic of an optically thick accretion disk. The H-alpha emission 
line radial velocity lags the expected position for the location of the 
accretion disk at phase 0.4-0.5. Most of the radiation appears to emanate 
from the opposite side of the disk from the hot spot location. \citep{Hoard96} invoke an enhanced thickness line absorption region at the 
edge of the disk at phases 0.6 to 0.9.

\subsection{V425 Cas}

V425 Cas is a VY Scl-type nova-like variable viewed at low inclination 
\citep{Ritter03}. Its FUV flux level and absence of P Cygni profiles 
indicate that V425 Cas was in an intermediate brightness state when it was 
observed with the IUE telescope \citep{Szkody85}. The IUE FES optical 
magnitude at the time of the IUE observation (see below) was $V = 16$ 
while in its deepest low state V425 Cas has $V = 18$, thus confirming the 
intermediate state. Kato (2001) discovered large amplitude 
oscillations with a 2.65 day period which they attribute to suppression of a disk instability
by irradiation. If the variation were really due to a 
dwarf nova type instability, then this object could be transitional 
between a VY Scl system and an ultra-short outburst period dwarf nova. If 
that were the case, this would be the shortest recurrence time for dwarf nova 
outbursts ever seen in an H-rich CV citep{Hunger85}.

\subsection{HL Aqr}

HL Aqr (PHL227) is a UX UMa-type nova-like variable \citep{Downes95} 
with a highly variable optical spectrum and is also classified as a SW 
Sextantis member. It has a reddening value $E(B-V) = 0.05$, but the white 
dwarf mass and accretion rate are unknown. \citet{Rodr07a} 
estimate that HL Aqr has an orbital inclination in the range $19? < i < 
27?$, which is much lower than that of the emission-dominated, 
non-eclipsing SW Sex stars ($i \sim 60-70$). This suggests the possibility 
of many low-inclination nova-likes actually being SW Sex stars, but with a 
very different spectroscopic appearance as they show significant 
absorption rather than being emission line-dominated \citep{Rodr07}. HL Aqr is a virtual spectroscopic twin of V3885 Sgr (\cite{Haefner87} and references therein). It shows pronounced irregular 
variations and coherent rapid oscillations with a period of 19.6 s.

In order to constrain the synthetic spectral fitting and try to reduce the 
number of free parameters, we carried out a thorough search of the 
published literature for the most accurately known system parameters. This 
included the compilations in \citet{Ritter03} and the Goettingen CV 
Cat website as well as publications documented in the Astrophysics Data 
Service. The most critical parameter for the model fitting, the distance, 
is one of the least known. Unlike dwarf novae which reveal a correlation 
between their absolute magnitude at maximum and their orbital period, 
there is no such relation for the nova-likes. However, a new method 
\citep{Knigge06} utilizing 2MASS JHK photometry and the observed properties 
of CV donor stars has proven useful for constraining nova-like distances. 
At present, this is the only reliable handle one has on nova-like 
distances although caution should be exercised because for 
some nova-like systems (e.g. TT Ari), the method appears to break down. 
For each system, we obtained the J,H,K apparent magnitudes from 2MASS 
photometry. For a given orbital period, \citet{Knigge06} provides absolute J, 
H and K magnitudes based upon his semi-empirical donor sequence for CVs. 
If it is assumed that the donor provides 100\% of the light in J, H and K, 
then the distance is a strict lower limit. If the donor emits 33\% of the 
light (the remainder being accretion light), then an approximate upper 
limit is obtained. The latter limit is a factor of 1.75 times the lower 
limit distance. The adopted distances used as constraints in the synthetic 
spectral fitting procedure are given in Table 1, where we list the adopted 
parameters for the orbital period (hours), the apparent V-magnitude, the 
inclination {\it{i}}($\degr$), the white dwarf mass (M$_{\sun}$), the 
interstellar reddening, E($\bv$), and the distance or distance range in 
parsecs.

\begin{deluxetable}{lcccccc}
\tablecaption{Nova-Like System Parameters}
\tablenum{1}
\tablecolumns{8}
\tablehead{
\colhead{System}
&\colhead{P$_{orb}$ (hours)}
&\colhead{V - Range}
&\colhead{$\it{i}$($\degr$)}
&\colhead{M$_{wd}$ (M$_{\sun}$)}
&\colhead{E ($\bv$)}
&\colhead{d$_{knigge}$ (pc)}
}
\startdata
KR Aur & 3.907 & $11.3 -16.9$  & $38\pm10$ &  $0.59\pm0.17$ & 0.05 & $779-1362$  \\ 
RW Tri & 5.57 &   $12.5 -15.6$ &  $70.5\pm2.5$ &  $0.55\pm0.15$ & 0.10 & $204-357$ \\  
V825 Her & 4.944 & 14.1 -  & - &  - &  -  & $380-665$ \\
V795 Her & 2.598 & $12.7 - 16.9$ & -  & -  & - & $115-202$ \\
BP Lyn  &    3.741 & $14.5 - 17.2$ & $79.8\pm5$ & - & - & $251-440$ \\
V425 Cas &  3.590 & $14.5 - 18$ & $25 \pm9$ & $0.86 \pm0.32$ & - &  $282-494$ \\  
HL Aqr & 3.25 &  $13.4 - 13.6$ & 18 & - & 0.05 & $174-304$ \\
\enddata
\end{deluxetable}

\section{Far Ultraviolet Spectroscopic Observations}

All the spectral data obtained from the Multimission Archive at Space 
Telescope (MAST) IUE archive are in a high activity state, very near or at 
outburst.  We restricted our selection to those systems with SWP spectra, 
with resolution of 5\AA\ and a spectral range of 1170\AA\ to 2000\AA.  
All spectra were taken through the large aperture at low dispersion.  
When more than one spectrum with adequate signal-to-noise ratio was 
available, the spectra were co-added or the two best spectra were 
analyzed.  In Table 2, an observing log of the IUE archival spectra is 
presented in which, by column: (1) lists the SWP spectrum number, (2) the 
aperture diameter, (3) the exposure time in seconds, (4) the date and time 
of the observation, (5) the continuum to background counts, and (6) the 
brightness state of the system.  Transition refers to an intermediate 
state between the highest optical brightness state and the deepest low 
state.

\begin{deluxetable}{lccccccc}  
\tablecaption{IUE Observing Log}
\tablenum{2}
\tablewidth{0pc}
\tablecolumns{7}
\tablehead{
\colhead{System}                                   
&\colhead{SWP}
&\colhead{t$_{exp}$}
&\colhead{Disp.}
&\colhead{Ap.}
&\multicolumn{2}{c}{Date of Observation}
&\colhead{State}
}
\startdata        
KR Aur & 13584 & 1920  & LOW & Lg  &\multicolumn{2}{c}{1981-3-26}  &  High   \\  
RW Tri & 07915 & 9000 & LOW & Lg  &\multicolumn{2}{c}{1980-2-11}  &  High   \\
RW Tri & 10135 & 7500 & LOW & Lg  &\multicolumn{2}{c}{1980-9-15}  &  High   \\
RW Tri & 16037 & 5400 & LOW & Lg  &\multicolumn{2}{c}{1982-1-13}  &  High   \\
RW Tri & 16041 & 5000 & LOW & Lg  &\multicolumn{2}{c}{1982-1-14}  &  High   \\
RW Tri & 16064 & 7200 & LOW & Lg  &\multicolumn{2}{c}{1982-1-18}  &  High   \\
RW Tri & 17617 & 3000 & LOW & Lg  &\multicolumn{2}{c}{1982-8-7}   &  High   \\
RW Tri & 17621 & 3500 & LOW & Lg  &\multicolumn{2}{c}{1982-8-7}   &  High   \\
V825 Her & 17353 & 2400  & LOW  & Lg   &\multicolumn{2}{c}{1982-7-04} &  High \\
V795 Her &  45334 & 3600 & LOW & Lg &\multicolumn{2}{c}{1992-8-13} & High   \\
BP Lyn & 32940 & 7200 & LOW  & Lg &\multicolumn{2}{c}{1988-2-18}  &  High   \\
V425 Cas & 14735 & 6900 & LOW & Lg &\multicolumn{2}{c}{1981-8-12} & Intermediate  \\
HL Aqr  & 23325  & 3600 & LOW & Lg & \multicolumn{2}{c}{1984-6-24} & High \\
\hline
\enddata
\end{deluxetable}

The activity state corresponding to each spectrum was determined by 
examining the AAVSO light curves as well as the flux level of the IUE 
spectrum that typically made it obvious the spectrum was obtained in a 
high state or intermediate state. In the case of those systems not covered 
by the AAVSO, their activity state was assessed based upon either mean 
photometric magnitudes taken from the \citet{Ritter03} catalogue or 
from IUE Fine Error Sensor (FES)  measurements at the time of the IUE 
observation. The FES counts, when available, can be converted to optical 
magnitudes to help ascertain the brightness state at the time of the IUE 
observation. In addition, the presence of P-Cygni profiles, absorption 
lines, and a comparison with spectral data and flux levels of other 
systems during different activity states was also used to ascertain or 
confirm the brightness state of the system. The reddening of the systems 
was taken from estimates listed in the literature, usually determined from 
the strength of the 2200\AA\ interstellar absorption feature if present. 
The three principal sources of reddening were the compilations of \citet{Verbunt87}, \citet{LaDous91} and \citep{Bruch94}. The IUE spectra were 
de-reddened with the IUERDAF IDL routine UNRED.

\section{Synthetic Spectral Fitting Models}

We adopted model accretion disks from the optically thick disk model grid 
of \citet{Wade98}. In these accretion disk models, the innermost 
disk radius, R$_{in}$, is fixed at a fractional white dwarf radius of $x = 
R_{in}/R_{wd} = 1.05$. The outermost disk radius, R$_{out}$, was chosen so 
that T$_{eff}(R_{out})$ is near 10,000K since disk annuli beyond this 
point, which are cooler zones with larger radii, would provide only a very 
small contribution to the mid and far UV disk flux, particularly the SWP 
FUV bandpass. The mass transfer rate is assumed to be the same for all 
radii.

Theoretical, high gravity, photospheric spectra were computed by first 
using the code TLUSTY \citep{Hubeny88} to calculate the atmospheric structure 
and SYNSPEC \citep{Hubeny95} to construct synthetic spectra. We 
compiled a library of photospheric spectra covering the temperature range 
from 15,000K to 70,000K in increments of 1000 K, and a surface gravity 
range, log $g = 7.0 - 9.0$, in increments of 0.2 in log $g$.

After masking emission lines in the spectra, we determined separately for 
each spectrum, the best-fitting white dwarf-only models and the 
best-fitting disk-only models using IUEFIT, a $\chi^{2}$ minimization 
routine. A $\chi^{2}$ value and a scale factor were computed for each 
model fit. The scale factor, $S$, normalized to a kiloparsec and solar 
radius, can be related to the white dwarf radius R through:  
$F_{\lambda(obs)} = S H_{\lambda(model)}$, where $S=4\pi R^{2} d^{-2}$, 
and $d$ is the distance to the source. For the white dwarf radii, we use 
the mass-radius relation from the evolutionary model grid of Wood (1995) 
for C-O cores. We combined white dwarf models and accretion disk models 
using a $\chi^{2}$ minimization routine called DISKFIT. Using this method 
the best-fitting composite white dwarf plus disk model is determined on 
the basis of the minimum $\chi^{2}$ value achieved, visual inspection of 
the model consistency with the continuum slope and Lyman Alpha region, and 
consistency of the scale factor-derived distance with the adopted Knigge 
(2006) distance for each system. Based upon formal error analyses carried 
out on synthetic spectral fitting of IUE spectra having comparable 
quality to the spectra studied here (e.g. \citet{Winter03}), we estimate 
that our accretion rates are accurate to within a factor of two to three.

\subsection{KR Aur}

Its two far UV spectra are dominated by absorption features due to metals 
with the strongest features due to N V (1240), Si III (1300), C II (1335), 
and Si IV (1400). Of the two SWP spectra, SWP 13584 has a slightly higher 
flux level and a better signal to noise. At the time the IUE spectra were 
taken, the system had an FES visual magnitude of 13.3 which is close 
to its most typical optical brightness of 13.5 but fainter than its 
highest brightness state at 12.5. There are no P Cygni profiles  
indicating wind outflow. We find that neither an accretion disk alone nor 
a white dwarf photosphere alone provides a satisfactory fit to the 
observed continuum slope and absorption spectrum of KR Aur. However, we 
found that a statistically significant improvement (lower $\chi^{2}$) in 
the fitting was evident when we combined an accretion disk model with a 
white dwarf model. We find that the optimal combination for KR Aur 
consists of an accretion disk model with a white dwarf mass of 0.55 
M$_{\sun}$, an inclination angle $i = 41$ degrees and an accretion rate of 
$3\times10^{-10}$M$_{\sun}$/yr together with a white dwarf photosphere 
model with T$_{eff} = 29,000$K$\pm 2000$K, log $g = 8$, for a distance of 
204 pc. This best combination fit to the IUE spectrum of KR Aur is  
displayed in Fig. 1. In this fit, the accretion disk provides 82\% of the 
FUV flux and the white dwarf contributes 18\% of the flux.

\subsection{RW Tri}

Between February 1980 and August 1982, twelve far ultraviolet SWP spectra 
were obtained with the IUE. These spectra can be characterized by 
prominent emission lines of Lyman Alpha, N V (1240), Si IV (1400), and C 
IV (1550) with weaker emission components at C III (1175), O III (1590), 
and He II (1640). However, absorption features are also seen at Si II 
(1264), Si III (1303), Si IV (1393,1402), C II (1721) and unidentified 
absorption features at 1655\AA, 1670\AA, and 1710\AA. The continuum flux 
level among the twelve spectra remains fairly constant. However, some of 
the spectra are quite noisy and thus unusable for our analysis. In Table 
2, we list the IUE spectra suitable for our analysis, their exposure 
times, aperture size, dispersion, date of observation and the brightness 
state of the system at the time the IUE spectrum was obtained. On the 
basis of the seven most suitable IUE spectra for model analysis, we have 
found the best-fitting accretion disk models for the parallax distance of 
340 pc to have inclination angles ranging from 60 degrees to 75 degrees, 
and in all seven spectra, a corresponding white dwarf mass of 0.4 
M$_{\sun}$. The accretion rate of RW Tri, averaged over the seven spectra, 
is $6 \times 10^{-9}$ M$_{\sun}$/yr. The best-fitting solution to each IUE 
spectrum is displayed in Figs. 2a,b,c,d,e,f,g for spectra SWP07915, 
SWP10135, SWP16037, SWP16041, SWP16064, SWP17617, and SWP17621, 
respectively.

\subsection{V825 Her}

Fortunately, V825 Her has a usable IUE archival spectrum. This one
spectrum reveals a FUV continuum with a continuum slope typical of
nova-like variables in their high brightness state. There is very strong
absorption in the SiV (1393, 1402) resonance lines with the components of
the doublet blended into one deep absorption feature. In addition, the C
IV doublet is seen in P Cygni structure indicating wind outflow when the
IUE spectrum was obtained while C III (1175), N V (1238, 1240) are seen in
weak absorption. Possible absorption at He II (1640) is difficult to
discern from the noise level in the spectrum. We carried out accretion
disk model fitting to its archival IUE spectrum. The range of distances to
V825 Her (see Table 1) led us to adopt 500 pc.  For this distance, we
found that the best-fitting accretion disk model out of the entire Wade
and Hubeny disk grid has a low mass white dwarf (M$_{wd} =
0.35$M$_{\sun}$), inclination of 18 degrees and an accretion rate of
$3\times 10^{-9}$M$_{\sun}$/yr. This accretion disk fit is shown in figure
3.

\subsection{V795 Her}

The IUE line spectrum is dominated by absorption features at NV (1240), C
IV (1550), Si III + O I (1300), C II (1335), possible O V (1371)  Si IV
(1393, 1402; both components resolved), possible He II (1640), N IV
(1718), and Al III (1854, 1862). The IUE spectra have flux levels 25\%
lower than the flux level in HST FOS spectra of V795 Her. The white dwarf
mass, orbital inclination, reddening, and accretion rate are unknown.
Moreover, the spectra, to our knowledge, have never been analyzed with
actual disk and photosphere models. For our adopted distance of 159 pc, we
found the best-fitting accretion disk model to have M$_{wd} =
0.8$M$_{\sun}$, inclination $i = 41$ degrees, and \.{M}$ =
10^{-10}$M$_{\sun}$/yr. This best-fit disk model is displayed in Fig.4.

\subsection{BP Lyn}

The IUE line spectrum is dominated by a strong C III (1175) emission line
and markedly variable absorption features at NV (1240), Si III + O I
(1300), C II (1335), possible O V (1371) Si IV (1393, 1402; both
components resolved), C IV (1548, 1551) some spectra in P Cygni structure;
some spectra no C IV present), possible He II (1640), N IV (1718), and Al
III (1854, 1862). The IUE spectrum itself (SWP32940) has a signal to noise
ratio of about 3:1.  In Fig. 5, we display our best-fitting optically
thick steady state accretion disk model to BP Lyn's IUE spectrum and
derive an accretion rate of $10^{-8}$M$_{\sun}$/yr for our distance of 344
pc.

\subsection{V425 Cas}

The low flux level is consistent with an intermediate to low brightness 
state and definable continuum but the signal to noise is insufficient to 
identify any lines with certainty. There is a strong emission feature at 
1660A and a possible emission feature at C IV (1550) but no  line 
features are clearly identifiable. With our adopted distance of 278 pc, we 
find that the best-fitting accretion disk model corresponds to M$_{wd} = 
0.8$ M$_{\sun}$, $i = 75$ degrees, and an accretion rate \.{M}$ = 1\times 
10^{-10}$M$_{\sun}$. This best-fitting solution is displayed in 
Fig.6. Szkody (1990) using \citet{Williams82a} 
models found \.{M}$ = 1\times 10^{-9}$M$_{\sun}$/yr while the Patterson 
(1984) P$_{orb}$ versus \.{M} relation yields $5\times 
10^{-10}$M$_{\sun}$/yr. Using the H$\beta$ equivalent widths and 
Patterson's relation, \.{M}$ = 10^{-10}$ M$_{\sun}$/yr, which is in 
agreement with our FUV-derived value.

\subsection{HL Aqr}

The FUV spectrum of this UX UMa system is characteristic of the UX UMa 
systems viewed at low orbital inclination. Strong wind absorption features 
of C IV (1550; in P Cygni structure), Si IV (1400) and N V (1240) are seen 
along with absorption due to C III (1175), Si III + O I (1300), C II 
(1335) and He II (1640). For the \citet{Knigge06} distance of 213 pc, our 
best-fitting disk model, shown in Fig.7, indicates a low mass white dwarf 
($\sim0.4$M$_{\sun}$), a low disk inclination angle of 18 degrees and an 
accretion rate of $1\times 10^{-9}$M$_{\sun}$/yr.

In Table 3, we summarize the best-fitting parameters of this selected 
sample of nova-like variables where the entries by column are (1) the 
system name, (2) nova-like subclass, (3) white dwarf mass, (4) inclination 
angle, (5) the distance in pc, (6) best-fitting model distance in pc, (7) 
\.{M} (M$_{\sun}$ yr$^{-1}$), and (6) 
minimum $\chi^{2}$ value.

\begin{deluxetable}{lccccc}
\tablecaption{Nova-Like System Parameters}
\tablenum{3}
\tablecolumns{8}
\tablewidth{0pc}
\tablehead{
\colhead{System}
&\colhead{M$_{wd}$ (M$_{\sun}$)}
&\colhead{\it{i}($\degr$)}
&\colhead{d$_{model}$ (pc)}
&\colhead{\.{M}(M$_{\sun}$yr$^{-1}$)}
&\colhead{$\chi^{2}$}
}

\startdata
KR Aur   & 0.55 & 41 & 204  & $3\times10^{-10}$  & 7.65\\ 
RW Tri   &  0.35 & 75  & 341  & $6.3\times10^{-9}$  & - \\
V825 Her & 0.35 & 18 & 500 &  $3\times10^{-9}$ & 1.87 \\    
V795 Her & 0.80  & 41 & 159  & $1\times10^{-10}$ & 3.47 \\
BP Lyn   & 0.35 & 75 & 344 & $1\times10^{-8}$ & 2.81\\
V425 Cas & 0.80 & 75 & 278  & $1\times10^{-10}$  & 4.29\\
HL Aqr   & 0.35 & 18  & 213 & $1\times10^{-9}$& 4.99\\  
\hline
\enddata
\end{deluxetable}

\section{Conclusions}

For the VY Sculptoris nova-like KR Aur, we find that combinations of 
models consisting of a hot white dwarf and an optically thick accretion 
disk provides a significantly improved model fit to its FUV spectrum 
compared with fits utilizing an accretion disk alone or a white dwarf 
photosphere alone. The best-fitting model is a combination of a white 
dwarf with T$_{eff} = 29,000\pm2000$K, log $g = 8.4$, contributing 18\% of 
the FUV flux and an accretion disk with accretion rate \.{M}$ = 3 \times 
10^{-10}$M$_{\sun}$/yr at an inclination of 41 degrees, contributing 82\%. 
Our results are broadly consistent with the analysis of Puebla et al. 
(2007) who applied a multi-parametric optimization fitting method to KR 
Aur. They explored a  range of white dwarf masses 0.4 to 0.8 M$_{\sun}$, 
inclinations $i = 30-50$ and an adopted Patterson (1984) distance of 
$d = 180$ pc. They found an accretion rate 
\.{M}$ = 4.5 \times 10^{-10}$ M$_{\sun}$/yr which is 
atypically low for a VY Scl nova-like variable in its high state. Moreover, 
Puebla et al. (2007) found that KR Aur should have a significant flux 
contribution of a hot white dwarf in addition to an accretion disk. 
Our results are not inconsistent with their reported analysis.

Further, the distance implied by our best-fitting model solution is 
similar to the Patterson (1984) distance used by \citet{Puebla07}.
Thus, KR Aur represents a second case (TT Ari being the other) of a  
a nova-like variable whose distance using Knigge's method yields a 
large disagreement with existing distance estimates from other methods.
Taking the temperature of the white dwarf indicated by our solution at 
face value, then the white dwarf temperature (29,000K) would be cooler 
than other white dwarfs of known temperature in VY Scl systems. For 
example, the white dwarf in TT Ari has T$_{eff} = 39,000$K, in MV Lyra 
47,000K and in DW UMa, 49,000K. It is unfortunate that a FUSE spectrum of 
KR Aur was not obtained since this would provided a definitive check on 
the white dwarf temperature.

Our analysis of seven archival IUE spectra of the SW Sex-type nova-like RW 
Tri using its parallax distance consistently yields a low mass ($\sim 
0.4$M$_{\sun}$) white dwarf and an average accretion rate, \.{M}$ = 
6.3\times 10^{-9}$ M$_{\sun}$/yr.  On the other hand, \citet{Puebla07}, 
using their statistical fitting solution, constrained the white dwarf mass 
to be in the range 0.4 to 0.8 M$_{\sun}$, and the inclination angle 
between 60 and 80 degrees for the fixed distance of 340 pc. Our modeling 
strongly favors a low mass white dwarf ( M$_{wd} < 0.6$ M$_{\sun}$ ). Our 
low white dwarf mass also favors the low end of the most likely mass range 
(0.4 to 0.7 M$_{\sun}$) estimated by \citet{Poole03}. A comparison of our 
accretion rate with other derived accretion rates for RW Tri reveals it is 
lower than the value $10^{-8}$M$_{\sun}$/yr determined by \citet{Groot04} but agrees with the accretion rate of $4.7 \times 
10^{-9}$M$_{\sun}$/yr derived independently by \citet{Puebla07}. 
However, our accretion rate is a factor of two larger than the accretion 
rate published by \citet{Rutten92} from eclipse mapping and a factor 
of ten smaller than the accretion rate of \citet{Horne85}, also 
derived from eclipse mapping.

Our model analysis of the UX UMa system V825 Her reveals an accretion rate 
of $3\times 10^{-9}$M$_{\sun}$/yr with a low inclination angle (18 
degrees) and a low white dwarf mass (M$_wd < 0.6$ M$_{\sun}$). For UX UMa 
systems which are stuck in permanent outburst, this high accretion rate is 
not unreasonable. Likewise for HL Aqr, another UX UMa systems, our best 
fit model indicates a low mass white dwarf, a low inclination angle but 
with an accretion rate $\times 10^{-9}$M$_{\sun}$/yr, a factor of 3 lower 
than V825 Her. Since HL Aqr is considered a spectroscopic twin of V3885 
Sgr, we note that a likely FUSE + HST STIS detection of the hot accreting 
white dwarf with T$_{eff} = 57,000$K in the UX UMa system, V3885 Sgr, by 
\citet{Linnell09} raises the possibility of an object with 
similar surface temperature being present in the UX UMa system HL Aqr.

For BP Lyn we find a high accretion \.{M}$ = 1 \times 
10^{-8}$M$_{\sun}$/yr. \citet{Hoard96} commented that the continuum 
slope is consistent with that of an accretion disk-only although no 
detailed accretion disk model with vertical structure was applied to the 
data. It is clear from \citet{Hoard96} that BP Lyn is a bona fide 
SW Sex member and appears to be a low inclination analog of RW Tri \citep{Still95}.

Finally, for V425 Cas, based upon a single IUE spectrum of this VY Scl
system taken during an intermediate brightness state, we find that an
accretion disk dominates its far UV spectrum but with a non-negligible
contribution from a hot white dwarf. Our derived accretion rate of \.{M}$
= 1\times 10^{-10}$M$_{\sun}$/yr is lower than typical accretion rates of
nova-likes during high brightness states which appears consistent with
V425 Cas being in an intermediate brightness state. Szkody (1990)  using
\citet{Williams82} models found \.{M}$ = 1\times
10^{-9}$M$_{\sun}$/yr while the \citet{Patterson84} P$_{orb}$ versus \.{M}
relation yields $5\times 10^{-10}$M$_{\sun}$/yr. Using the H$\beta$
equivalent widths and Patterson's relation, \.{M}$ = 10^{-10}$
M$_{\sun}$/yr, this is in agreement with our FUV-derived value.

\citet{Ballouz09} applied multi-component modeling to the archival 
FUV spectra of nova-like variables that are members of the SW Sextantis 
subclass and those which are not classified as members. They found no 
difference between the average derived accretion rates of SW Sex members 
and non-SW Sex members. The present study adds the modeling of seven 
nova-like systems to the sample of 15 nova-like systems analyzed by 
\citet{Ballouz09}, 23 nova-likes analyzed by \citet{Puebla07}, 
and three nova-likes analyzed by \citet{Zellem09}. The nova-like 
systems in the present study represent a mix of different subclasses (UX 
UMa-type, VY Scl-type and SW Sex-type) whose archival FUV spectra have 
been compared, typically for the first time, with reasonably realistic 
models of accretion disks and photospheres. All but one had FUV spectra 
taken in their high brightness states.
  
The accretion rates we have derived in this paper are only as good as the 
distances we have adopted. The method of \citet{Knigge06} represents the best 
handle we presently have on the distances to nova-like variables. Since any 
derived accretion rates depend very sensitively upon distance, precision 
trigonometric parallaxes from the ground and from space (e.g. GAIA) are required 
to determine the most accurate accretion rates for the different subtypes of nova-like 
variables. In the meantime, more nova-like parallaxes are needed from the 
ground and with the fine guidance sensor (FGS) on HST.

We thank an anonymous referee for useful comments.
This work is supported by NSF grant AST-0807892 to Villanova University 
and by the Delaware Space Grant Consortium.

\clearpage
\begin{figure}
\epsscale{0.8}
\plotone{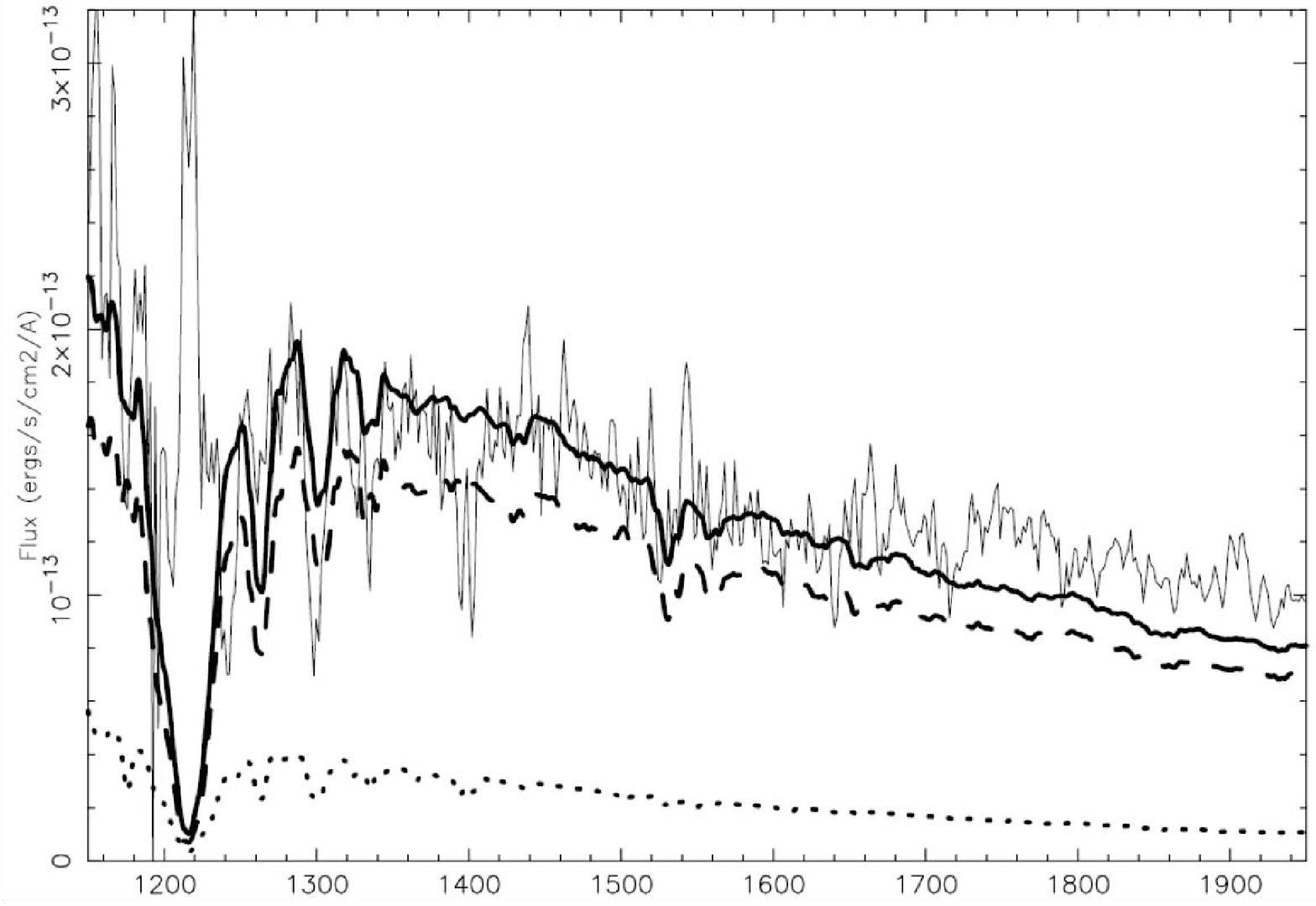}
\caption{The best-fitting combination accretion disk+white dwarf photosphere
synthetic fluxes to the spectrum SWP27096 of the VY Scl-type nova-like variable 
KR Aur during a high state. The accretion
disk corresponds to \.{M}$ = 3\times 10^{-10}$ M$_{\sun}$ yr$^{-1}$, ${i} = 41\degr$,
and M$_{wd} = 0.55$ M$_{\sun}$. The top solid curve is the
best-fitting combination, the dotted curve is the contribution of the white dwarf 
alone and the dashed curve is the accretion disk synthetic spectrum alone.
In this fit, the white dwarf with T$_{eff} = 29,000$K, Log $g = 8$ contributes 18\% of the FUV light
and the accretion disk contributes 82\% of the far UV flux.}
\end{figure}

\clearpage
\begin{figure}
\epsscale{0.8}
\plotone{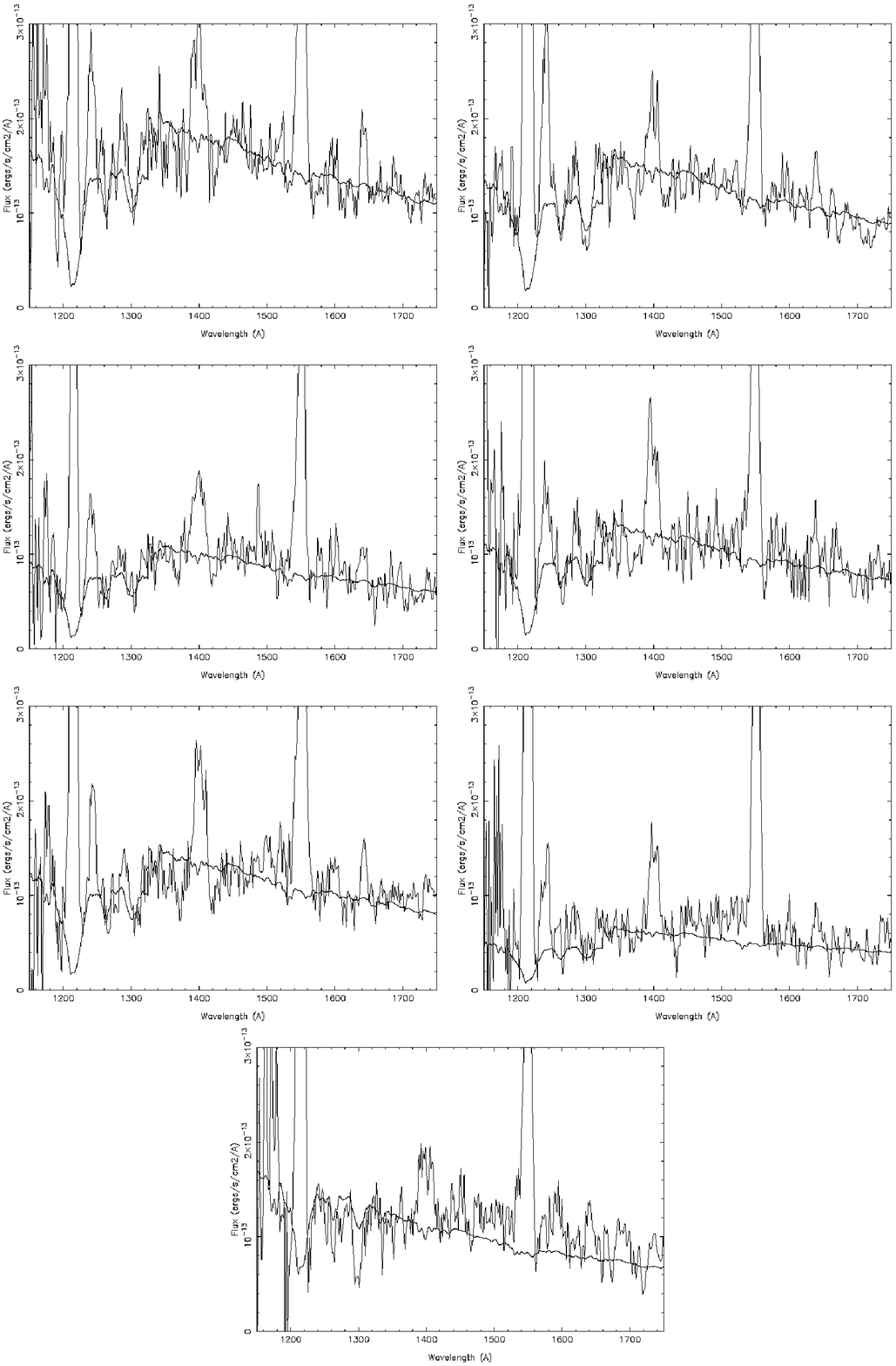}
\caption{The best-fitting accretion disk models to the IUE SWP spectra of 
the SW Sextantis-type nova-like system RW Tri during its high state. The 
seven panels from top to bottom are for spectra SWP07915, 
SWP10135, SWP16037, SWP16041, SWP16064, SWP17617, and SWP17621, 
respectively. The average accretion rate over 
the best disk model fits to the seven spectra corresponds to \.{M}$ = 
6.3\times 10^{-9}$ M$_{\sun}$ yr$^{-1}$, ${i} = 75\degr$, and M$_{wd} = 
0.4$ M$_{\sun}$.}
\end{figure}

\clearpage
\begin{figure}
\includegraphics[scale=0.75,angle=270]{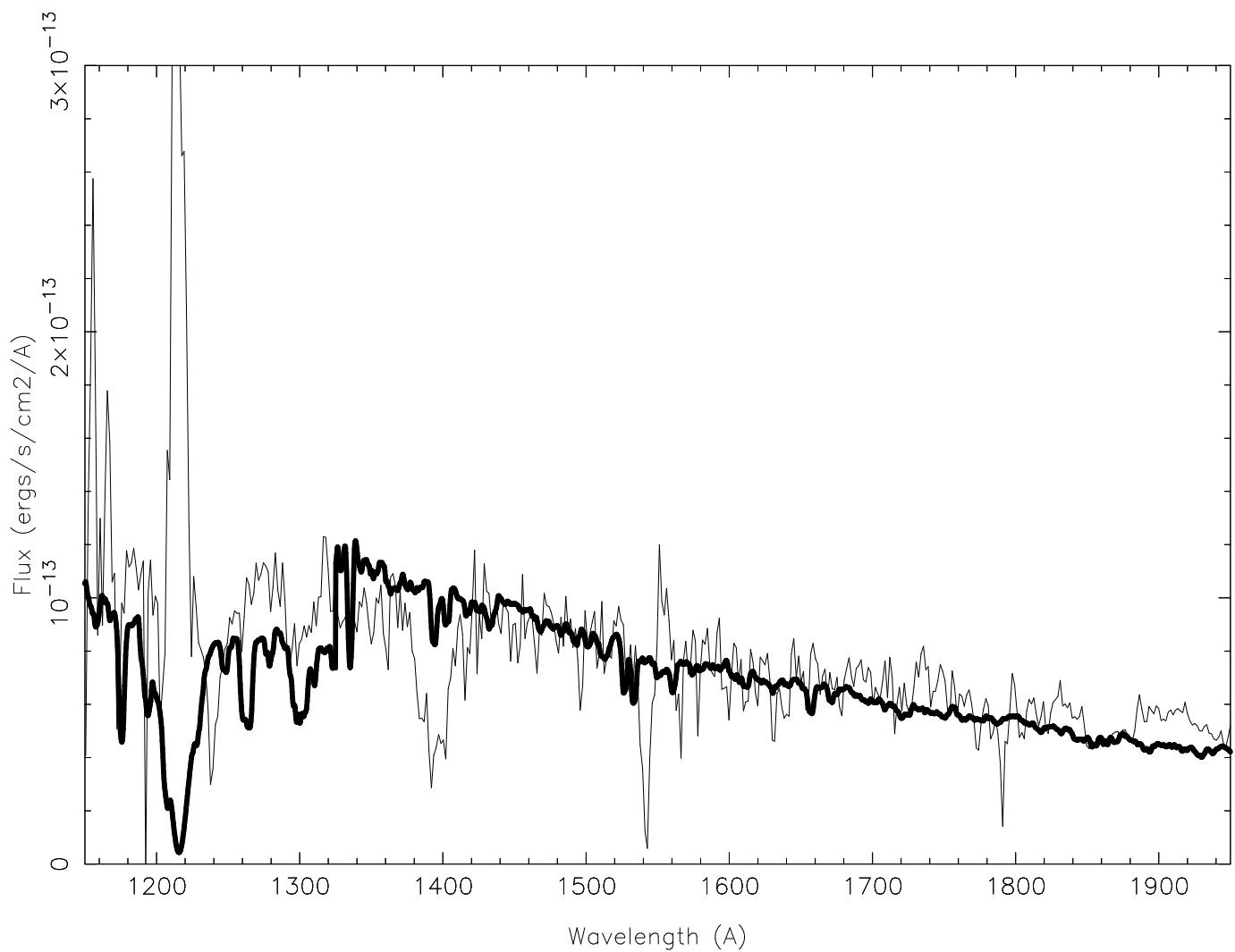}
\caption{The best-fitting accretion disk model to the IUE spectrum SWP17353 of the nova-like system V825 Her
during its high state. The accretion disk corresponds to \.{M}$ = 3.0\times 10^{-9}$ M$_{\sun}$ yr$^{-1}$, 
${i} = 18\degr$, and M$_{wd} = 0.4$ M$_{\sun}$ and a distance of 500 pc.   }
\end{figure}

\clearpage
\begin{figure}
\includegraphics[scale=0.75,angle=270]{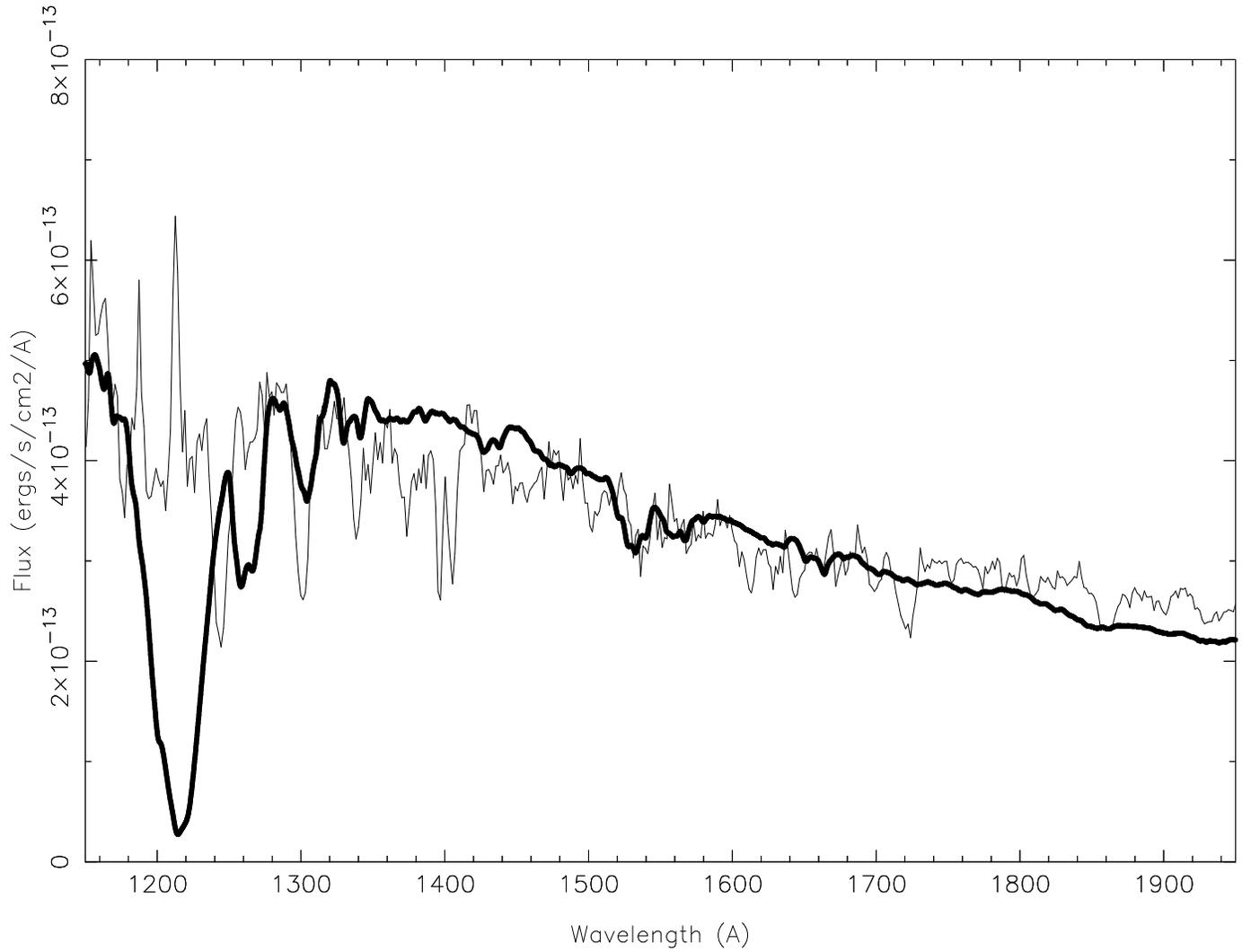}
\caption{The best-fitting accretion disk model to the IUE spectrum SWP45334 of the SW Sex-type nova-like system V795 Her
during its high state. The accretion disk corresponds to \.{M}$ = 1.0\times 10^{-10}$ M$_{\sun}$ yr$^{-1}$, 
${i} = 41\degr$, and M$_{wd} = 0.8$ M$_{\sun}$ and a distance of 159 pc.   }
\end{figure}

\clearpage
\begin{figure}
\includegraphics[scale=0.75,angle=270]{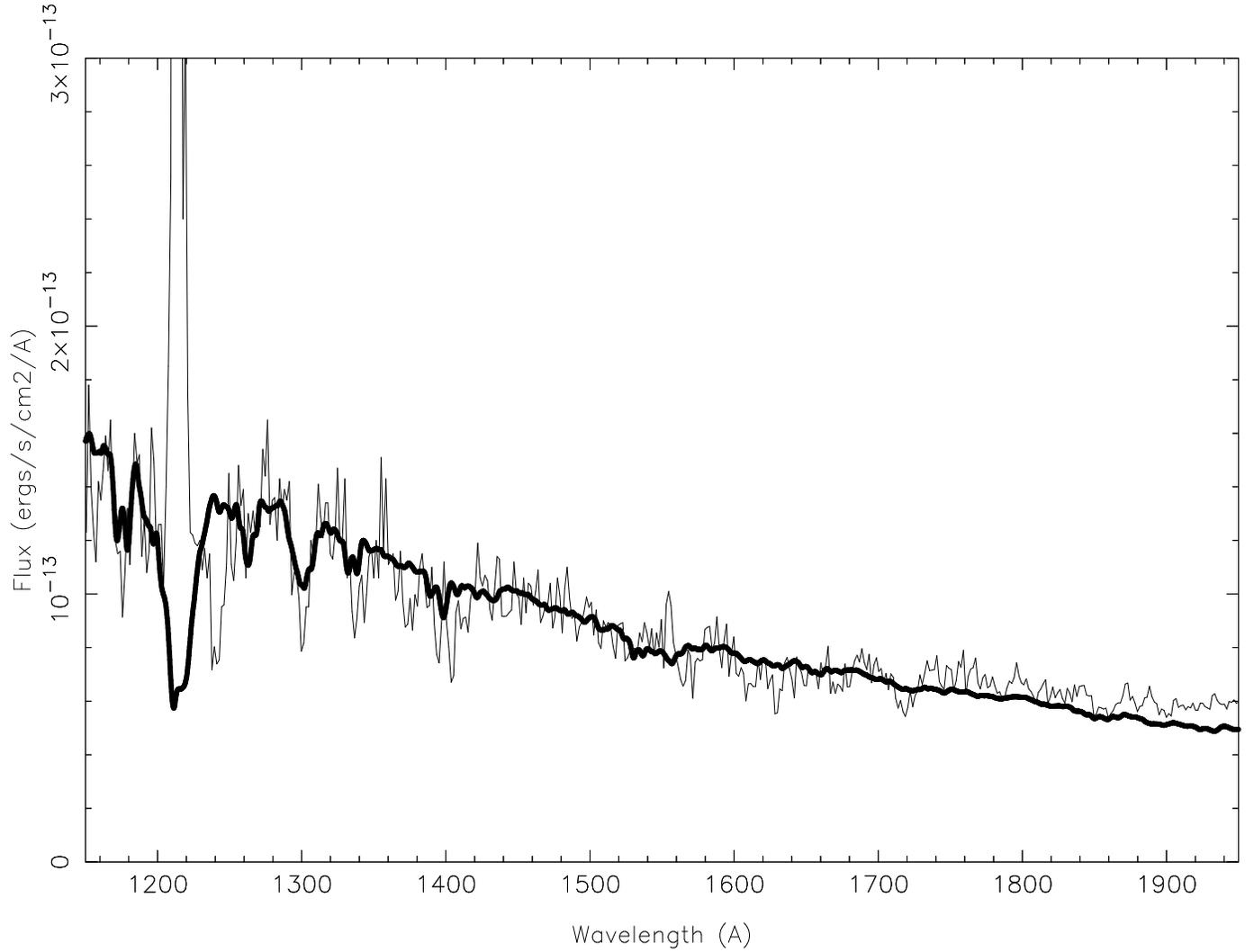}
\caption{The best-fitting accretion disk model to the IUE spectrum SWP32940 of the SW Sex-type (and a VY Scl-type) nova-like system BP Lyn
during its high state. The accretion disk corresponds to \.{M}$ = 1.0\times 10^{-8}$ M$_{\sun}$ yr$^{-1}$, 
${i} = 75\degr$, and M$_{wd} = 0.4$ M$_{\sun}$ and a distance of 344 
pc.}
\end{figure}

\clearpage
\begin{figure}
\includegraphics[scale=0.75,angle=270]{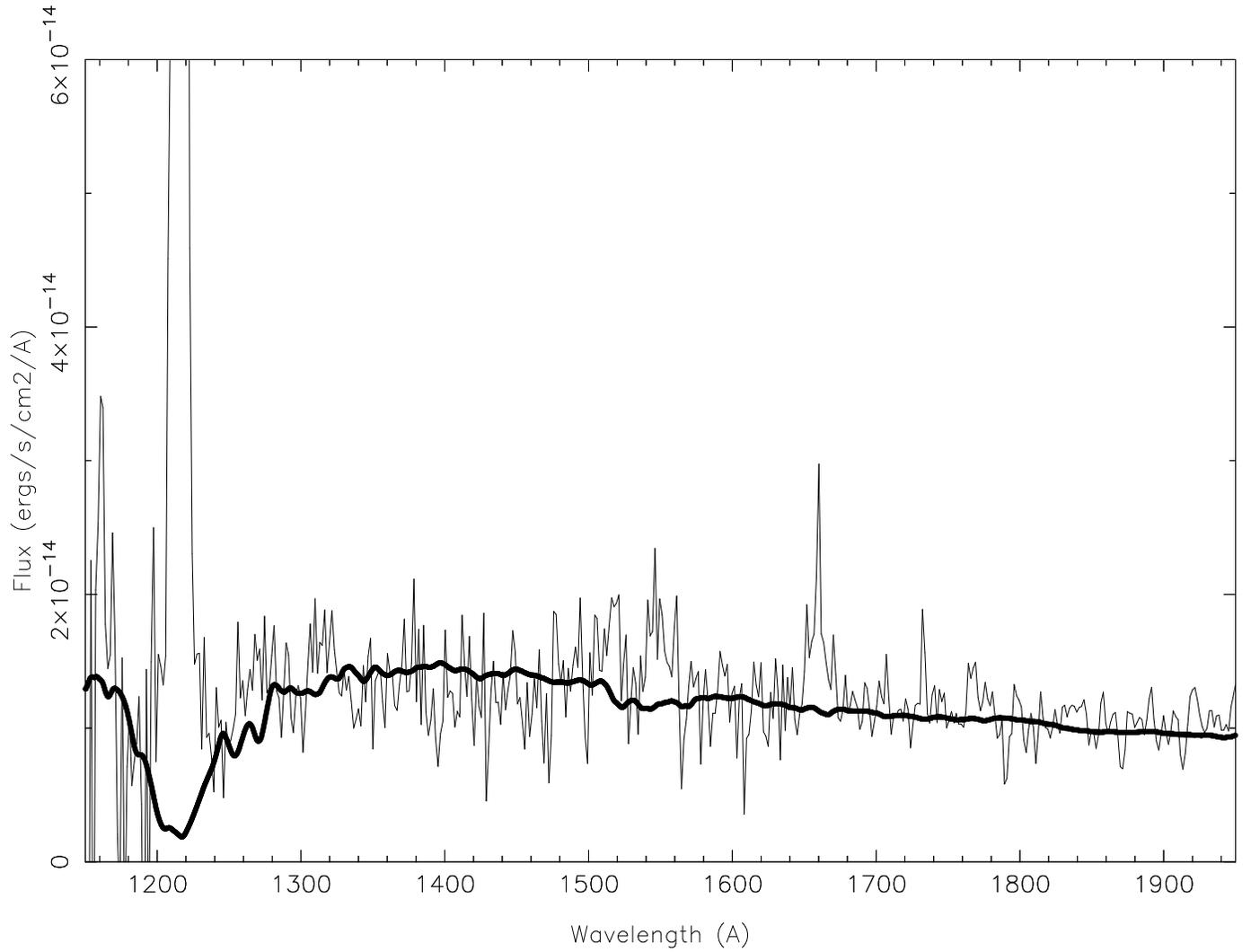}
\caption{The best-fitting accretion disk model to the IUE spectrum SWP14735 of the VY Scl-type nova-like system V425 Cas
during its high state. The accretion disk corresponds to \.{M}$ = 1.0\times 10^{-10}$ M$_{\sun}$ yr$^{-1}$, 
${i} = 75\degr$, and M$_{wd} = 0.8$ M$_{\sun}$ and a distance of 278 pc.}
\end{figure}

\clearpage
\begin{figure}
\includegraphics[scale=0.75,angle=270]{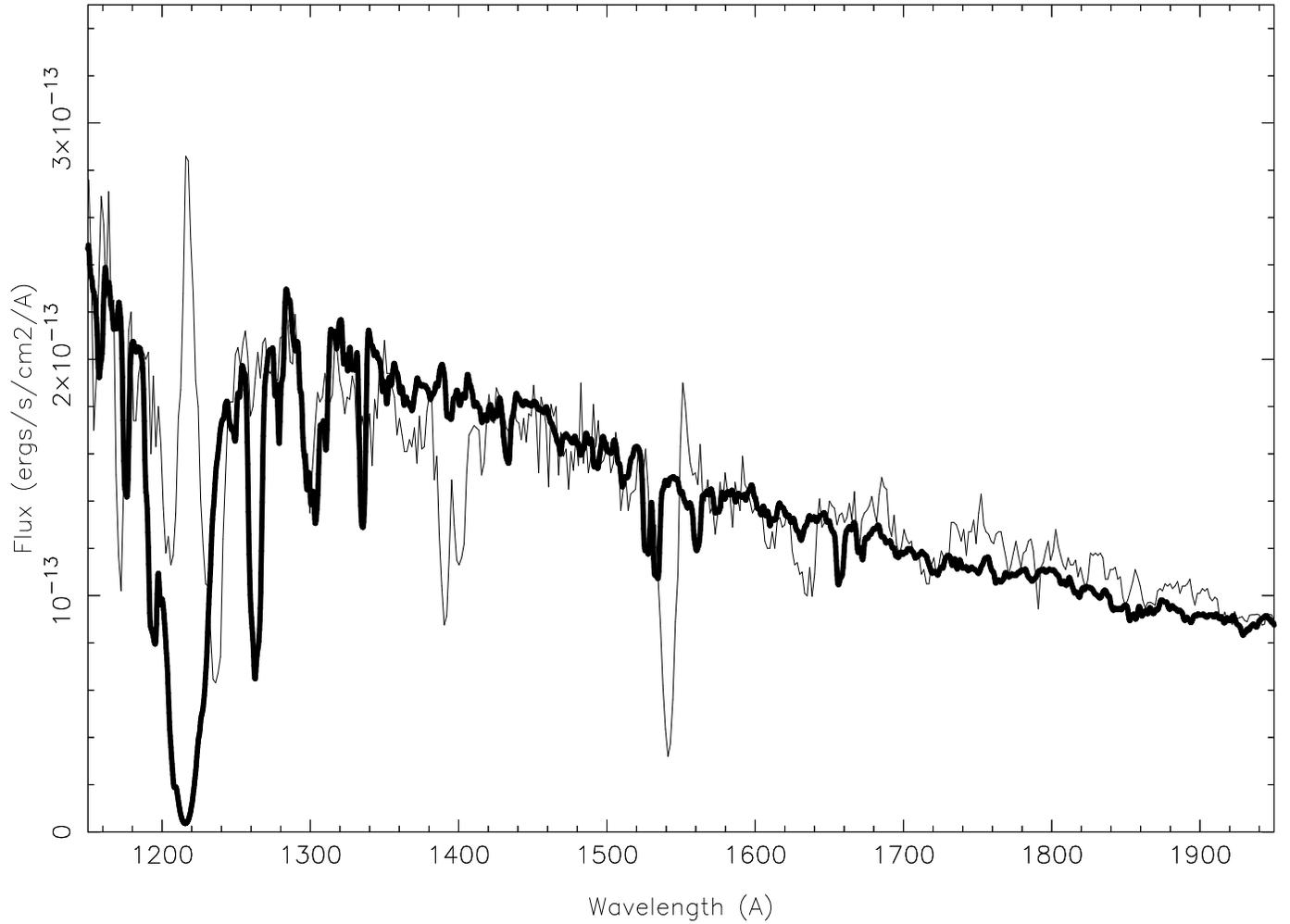}
\caption{The best-fitting accretion disk model to the IUE spectrum SWP23325 of the SW Sex-type nova-like system HL Aqr
during its high state. The accretion disk corresponds to \.{M}$ = 1.0\times 10^{-9}$ M$_{\sun}$ yr$^{-1}$, 
${i} = 18\degr$, and M$_{wd} = 0.4$ M$_{\sun}$ and a distance of 213 pc.}
\end{figure}

\end{document}